\documentclass[pra,aps,twocolumn,showpacs,10pt]{revtex4-1}
\usepackage{amsmath,amssymb,graphicx,amsthm,dsfont,bm,color,ulem}

\begin{document}

\title{An upper limit to the orbital angular momentum of a vortex-carrying ultrashort pulse}
%\title{An upper bound to the topological charge of the vortex carried by a ring pulse}

\author{Miguel A. Porras}

\affiliation{Grupo de Sistemas Complejos, ETSIME, Universidad Polit\'ecnica de Madrid, Rios Rosas 21, 28003 Madrid, Spain}

\begin{abstract}
The magnitude of the topological charge of the vortex of a ring-shaped pulse is found to be limited by the squared mean frequency of the pulse spectrum at the ring relative to its variance. This limitation implies a upper bound to the orbital angular momentum carried by a pulse, and a lower bound to the duration for a pulse to carry any orbital angular momentum.
\end{abstract}

%\pacs{32.80.Wr; 42.65.Tg; 05.45.Yv}

\maketitle

\noindent An optical vortex beam is an optical beam with a phase variation $e^{-il\phi}$ in the azimuthal direction $\phi$ perpendicular to the propagation direction of the beam, say $z$. At the beam center, $r=0$, the phase is undetermined and the optical field strength vanishes, which endows the beam with a ring structure. The integer number $l$, called topological charge, can take arbitrarily high positive or negative values and determines the orbital angular momentum (OAM) $l\hbar$ per photon of the beam, in addition to its spin angular momentum $\pm\hbar$ \cite{ALLEN,YAOPADGETT}. Archetypical vortex beams are Laguerre-Gauss (LG) beams, easily produced from a vortex-less beam with a fork-type grating or a spiral phase plate \cite{YAOPADGETT}. Recognizing the existence of these new and unlimited degrees of freedom of the angular momentum of the photons in a beam as familiar as a LG beam \cite{ALLEN} has led to a revolution in optics with ramifications in other fields such as astrophysics \cite{FOO} or medicine \cite{JEFFRIES}, with applications as optical tweezers \cite{HE}, to laser ablation \cite{OMATSU} or classical and quantum information processing \cite{GIBSON,MAIR}. The unboundedness of $l$ seems to be corroborated by the feasibility of generating beams with up to $10^4\hbar$ OAM per photon \cite{MATSKO}.

With ultrashort pulses these applications acquire, in addition, ultrafast resolution. Generation of shorter and shorter ring-shaped pulses carrying vortices has had to overcome practical difficulties such as the spatial, group velocity and topological charge dispersion introduced by fork gratings or spiral phase plates \cite{MARIYENKO,ZEYLIKOVICH,TOKIZANE}. Nonetheless the few-optical-cycle regime has already been achieved \cite{SHVEDOV,YAMANE}, and in strong-field physics \cite{ZURCH} these few-cycle, vortex-carrying pulses are used to excite extremely high harmonics and extremely short, attosecond pulses characterized by high topological charges \cite{ZURCH,HERNANDEZ,GARIEPY,REGO}.

In this Letter we show that there is a fundamental restriction to the topological charge of the vortex, and hence to the OAM, that a ring-shaped, ultrashort pulse can carry. As a reference, the topological charge of a single-cycle pulse, according to the standard definition in \cite{BRABEC}, is  $|l|=27$ as much. This restriction implies that there exist minimal $l$-vortex-carrying wave packets, and in particular, wave packets that cannot carry OAM at all. A limit to the OAM has been suggested to exist in ``single-cycle" (actually much shorter \cite{BRABEC}) X-waves with a particular frequency spectrum \cite{CONTI}, but X-waves have unbounded energy, and hence unbounded OAM \cite{ALLEN,YAOPADGETT}. Here a proof of a precise limit for general pulse shapes and for the actual beam geometries involved in experiments is provided.

Suppose we have overcome all the technical difficulties \cite{MARIYENKO,ZEYLIKOVICH,TOKIZANE} and synthesized the ring pulsed beam (RPB) 
\begin{equation}\label{AN}
E(r,\phi,z,t')= \frac{1}{\pi}\int_0^{\infty} \hat E_\omega(r,\phi,z)e^{-i\omega t'} d\omega\, ,
\end{equation}
superposition of LG beams
\begin{equation}\label{LG}
\hat E_\omega(r,\phi,z) = \hat a_\omega \frac{e^{-i(|l|+1)\psi_\omega(z)}e^{-il\phi}}{\sqrt{1+\left(\frac{z}{z_{R,\omega}}\right)^2}}\left(\frac{\sqrt{2}r}{s_\omega(z)}\right)^{|l|}e^{\frac{i\omega r^2}{2cq_\omega(z)}} \, ,
\end{equation}
all them of the same topological charge $l$, zero radial order, but different frequencies $\omega$ and weights $\hat a_\omega$.
In the above equations, and in the introduction, $(r,\phi,z)$ are cylindrical coordinates, $t'=t-z/c$ is the local time, $c$ is the speed of light in vacuum, $q_\omega(z)=z-iz_{R,\omega}$ is the complex beam parameter, $\psi_\omega(z)=\tan^{-1}(z/z_{R,\omega})$, $s_\omega(z)=s_\omega\sqrt{1+(z/z_{R,\omega})^2}$, and $s_\omega=\sqrt{2z_{R,\omega} c/\omega}$ is the waist width of the fundamental Gaussian beam ($l=0$). Additionally, the complex beam parameter is usually written as
\begin{equation}\label{Q}
\frac{1}{q_\omega(z)} = \frac{1}{R(z)} + i \frac{2c}{\omega s^2_\omega (z)}\, ,
\end{equation}
where $1/R_\omega(z)= z/(z^2+ z_{R,\omega}^2)$ is the curvature of the wave fronts. Being limited to positive frequencies, the optical field $E$ in (\ref{AN}) is the analytical complex representation of the real optical field $\mbox{Re}\{E\}$ \cite{BORN}. Since (\ref{AN}) vanishes at $r=0$ and $r\rightarrow\infty$, it will feature a bright ring about a certain radius. It is natural to identify, both from a theoretical and applicative point of view, the pulse shape of the RPB with that at the bright ring, or at the brightest if there are several. As said, we analyze if this pulse shape,
\begin{equation}\label{P}
P(t)= \frac{1}{\pi}\int_0^{\infty} \hat P_\omega e^{-i\omega t}d\omega\,,
\end{equation}
and the topological charge $l$ can be taken arbitrarily or there is some kind of restriction.

%We can then agree to identify the pulse shape of the VPB with the pulse shape at the ring, or at the brightest one if there are several, also because %in all applications in high-power and femtosecond or attosecond optics this pulse shape is the most relevant. The purpose of this Letter is to study %whether a VPB of certain pulse shape can carry arbitrary topological charge or there are some restrictions.

\begin{figure}
\includegraphics[width=7.5cm]{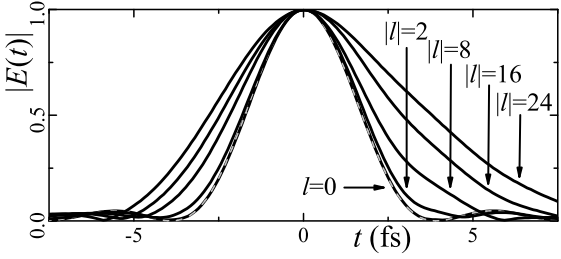}
\caption{\label{Fig1} Propagation of the pulsed LG disturbance  $E(r,\phi,0,t)=P(t)\left(\sqrt{2}r/s\right)^{|l|}\exp(-r^2/s^2)\exp(-il\phi)$ at $z=0$. Dashed gray curve: Envelope of the single-cycle pulse $P(t)=\mbox{sinc}^2(t/T)\exp(-i\omega_0 t)$, $\omega_0=2.417$ fs$^{-1}$, $T=3.9$ fs at $z=0$. Black curves: For the indicated values of $|l|$, envelope of the pulse at the bright maximum at one Rayleigh distance. Peak values are set to unity and shifted to $t=0$ for better comparison.}
\end{figure}

A prerequisite to talk about the pulse shape of a vortex-carrying pulse is that it remains unchanged during propagation (except for a global complex amplitude, as usually understood). Figure \ref{Fig1} illustrates the situation with the usual model with factorized field in space and time at the focus or waist $z=0$, in which the Gaussian waist width $s_\omega\equiv s$ is independent of frequency \cite{HERNANDEZ}. While for a few-cycle pulsed Gaussian beam ($l=0$) the pulse shape on its maximum (at $r=0$) is substantially unaltered \cite{PORRAS5,PORRAS6}, the same few-cycle pulse on the bright ring of a RPB widens and distorts, particularly for high $|l|$, as seen in Fig. \ref{Fig1}. The origin of this distortion is a dispersion, particularly enhanced for high $|l|$ and for short pulses, induced by Gouy's phase when $z_{R,\omega}$ depends on frequency, as it is in the factorized model. Similar dispersion affects at sufficiently high $|l|$ the pulse shape in any other model in which $z_{R,\omega}$ depends on frequency. Thus, the so-called isodiffracting model \cite{PORRAS1,PORRAS2,FENG,PORRAS3,PORRAS4} in which $z_{R,\omega}\equiv z_R$ is independent of frequency becomes particularly relevant as the only type of RPB for which an ultrashort pulse can maintain its shape during propagation irrespective of the value of $|l|$. With $\omega$-independent $z_R$, $\psi_\omega(z)\equiv \psi(z)$, $q_\omega(z)\equiv q(z)$ and $R_\omega(z)\equiv R(z)$ are also $\omega$-independent, and the Gaussian width $s_\omega(z)$ is inversely proportional to the square root $\omega$.

For our analysis we introduce the diffraction and azimuthal factor $D(z,\phi)=e^{-il\phi}e^{-(|l|+1)\psi(z)}/\sqrt{1+(z/z_R)^2}$ and the scaled radius $\rho=r/\sqrt{2z_R c[1+(z/z_R)^2]}$ at each distance $z$. Constant $\rho$ represents a hyperboloid of revolution about the $z$ axis, also called caustic surface, along which the ring pulse spreads. Using (\ref{Q}) and the expressions of $s_\omega(z)$ and $s_\omega$ above, integral in (\ref{AN}) with (\ref{LG}) is conveniently written as
\begin{equation}\label{ANLG}
E= D \left(\sqrt{2}\rho\right)^{|l|} \frac{1}{\pi}\int_0^\infty \hat a_\omega \omega^{|l|/2}e^{-\rho^2\omega}e^{-i\omega t^{\prime\prime}},
\end{equation}
where $t^{\prime\prime}=t'-r^2/2cR(z)$, and where it is seen that the pulse shape depends on the particular caustic surface $\rho$, but does not change on propagation, being only attenuated by the diffraction factor in $D$. If at $z=0$ the pulse peaks at the time $t'=0$, it does at $z\neq 0$  at the time determined by $t^{\prime\prime}\equiv t'-r^2/2cR(z)=0$, which defines a spherical pulse front of radius $R(z)$ at each distance $z$ matching the wave fronts of all monochromatic LG constituents.

We first find the expression of a RPB with the pulse shape $P(t)$ in (\ref{P}) at a particular caustic surface $\rho_p$. Equating (\ref{ANLG}) particularized at $\rho_p$ to $DP(t)$, we get $\hat a_\omega= e^{\rho_p^2\omega}\omega^{-|l|/2}(\sqrt{2}\rho_p)^{-|l|}\hat P_\omega$, and  (\ref{ANLG}) becomes
\begin{eqnarray}
E&=&D\left(\frac{\rho}{\rho_p}\right)^{|l|}\frac{1}{\pi}\int_0^{\infty} \hat P_\omega e^{-(\rho^2-\rho_p^2)\omega}e^{-i\omega t^{\prime\prime}} d\omega \nonumber \\
&=&D\left(\frac{\rho}{\rho_p}\right)^{|l|} P\left[t^{\prime\prime}- i (\rho^2-\rho_p^2) \right]\, . \label{ISO}
\end{eqnarray}
Its spectrum $\hat E_\omega = D(\rho/\rho_p)^{|l|} \hat P_\omega e^{-(\rho^2-\rho_p^2)\omega}$ at $\rho>\rho_p$ ($\rho<\rho_p$) is a red-shifted (blue-shifted) version of $\hat P_\omega$.

Next we identify the bright caustic surface as that where the pulse energy is maximum. The energy per unit transverse area, also called fluence, is given by ${\cal E}(r,z) =\int_{-\infty}^\infty (\mbox{Re} E)^2 dt = (1/2)\int_{-\infty}^\infty |E|^2 dt = (1/\pi)\int_0^\infty |\hat E_\omega|^2 d\omega$, and for the RPB in (\ref{ISO}) by
\begin{equation}\label{FLUENCE}
{\cal E} = |D|^2\left(\frac{\rho}{\rho_p}\right)^{2|l|}\frac{1}{\pi}\int_0^{\infty} |\hat P_\omega|^2 e^{-2(\rho^2-\rho_p^2)\omega}d\omega\, .
\end{equation}
Differentiating with respect to $\rho$, we obtain, after straightforward algebra, $d{\cal E}/d\rho = {\cal E}\left[2|l|- 4\rho^2\bar\omega(\rho)\right]/\rho$,
where
\begin{equation}\label{CARRIERR}
\bar\omega(\rho) = \frac{\int_0^\infty |\hat P_\omega|^2 e^{-2(\rho^2-\rho_p)\omega} \omega d\omega}{\int_0^\infty |\hat P_\omega|^2 e^{-2(\rho^2-\rho_p^2)\omega} d\omega}\,,
\end{equation}
is the mean frequency of the pulse at the caustic $\rho$. Thus, a caustic surface $\rho_s$ of maximum or minimum energy density satisfies $\rho_s^2 = |l|/2\bar \omega(\rho_s)$. The second derivative of the energy profile can be similarly evaluated and, at the maxima or minima $\rho_s$ satisfying $\rho_s^2 = |l|/2\bar \omega(\rho_s)$, is given by
$d^2{\cal E}/d\rho^2|_{\rho_s} = -8{\cal E}(\rho_s)\bar\omega(\rho_s)\left\{1-|l|[\sigma^2(\rho_s)/\bar\omega^2(\rho_s)]\right\}$,
where
\begin{equation}
\sigma^2(\rho) = \frac{\int_0^\infty |\hat P_\omega|^2 e^{-2\left(\rho^2-\rho_p^2\right)\omega} [\omega-\bar\omega(\rho)]^2 d\omega}{\int_0^\infty |\hat P_\omega|^2 e^{-2\left(-\rho^2-\rho_p^2\right)\omega} d\omega}\,
\end{equation}
is the variance of the pulse spectrum at each caustic $\rho$. Thus, the caustic $\rho_s$ has a maximum of energy if $|l|< \bar\omega^2(\rho_s)/\sigma^2(\rho_s)$, and a minimum if $|l| > \bar\omega^2(\rho_s)/\sigma^2(\rho_s)$.

Then, for $P(t)$ to be the pulse shape at a caustic or maximum or minimum energy, it must be located at $\rho_p^2=|l|/2\bar\omega$, where $\bar\omega\equiv\bar\omega(\rho_p)$ is the mean frequency of $P(t)$. The caustic is bright, in the sense of maximum pulse energy, if
\begin{equation}\label{MAXL}
|l| < \bar\omega^2/\sigma^2\, ,
\end{equation}
where $\sigma^2 \equiv \sigma^2(\rho_p)$ is the variance of the pulse spectrum, and a dark caustic or of minimum energy if $|l|>\bar\omega^2/\sigma^2$. In the latter case, for a continuous energy profile vanishing at $\rho=0$ and at $\rho\rightarrow \infty$, there must exist at least two maxima $\rho_s$ surrounding the minimum at $\rho_p$. In any of these maxima, e. g., the global maximum, the condition of maximum $|l|< \bar\omega^2(\rho_s)/\sigma^2(\rho_s)$ is satisfied by the pulse shape at that maximum. In conclusion, all RPBs verify restriction (\ref{MAXL}) between its topological charge and the pulse shape at its bright ring.

\begin{figure}[t!]
\centering
\includegraphics*[width=4.1cm]{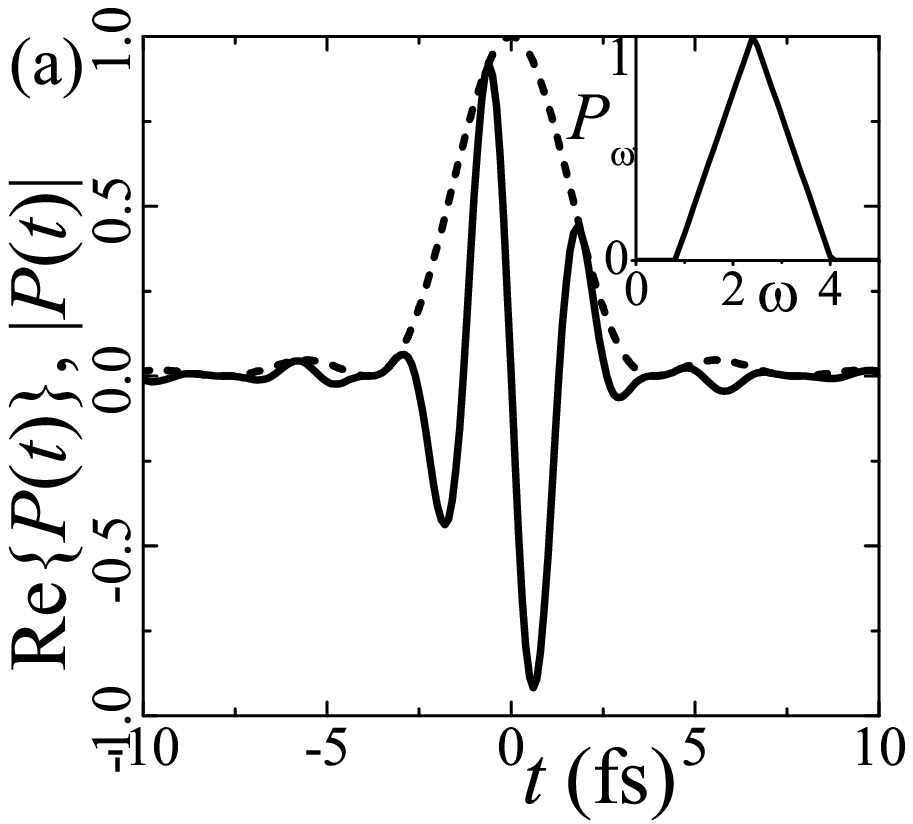}\includegraphics*[width=4.1cm]{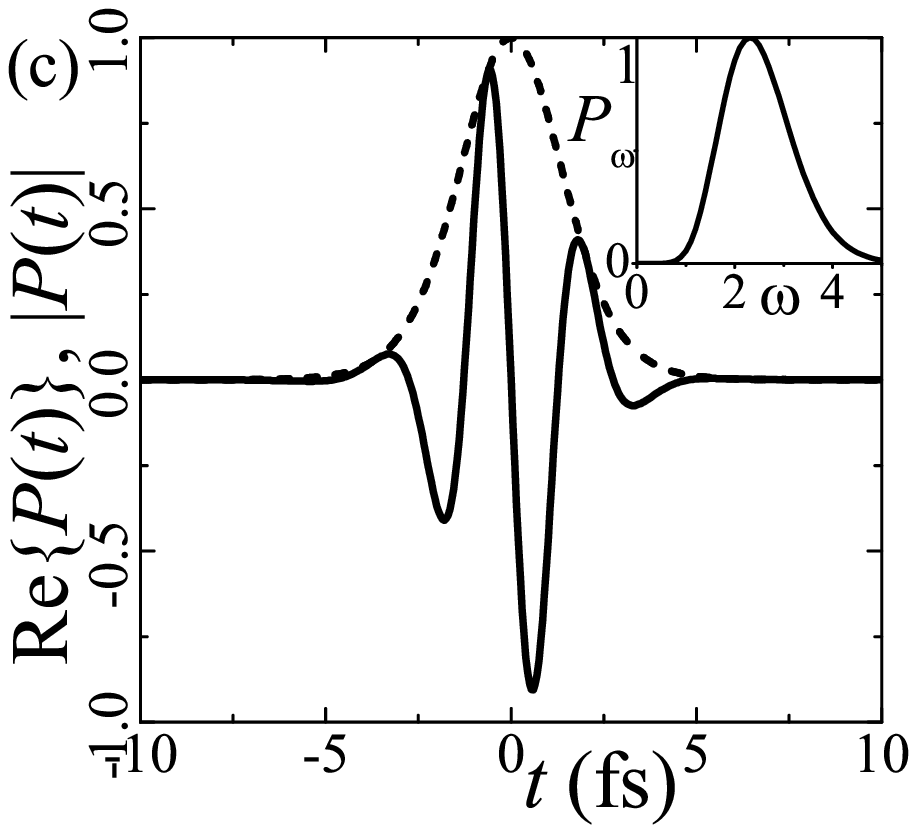}
\includegraphics*[width=7.8cm]{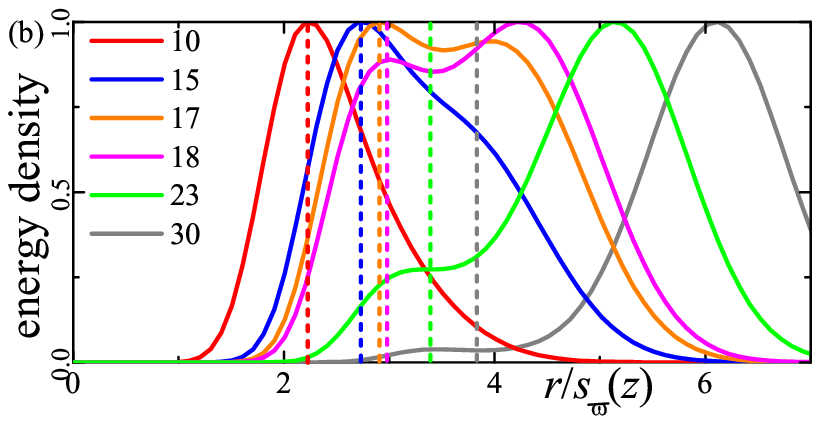}
\includegraphics*[width=7.8cm]{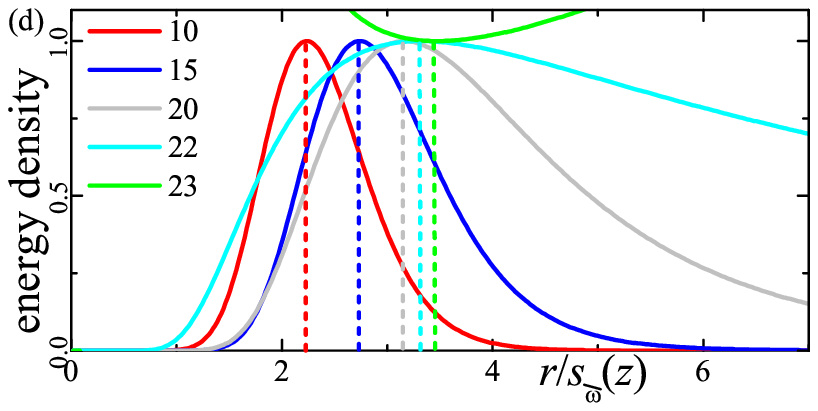}
\caption{\label{Fig2} (a) Real pulse and envelope of the complex analytical signal $P(t)=\mbox{sinc}^2(t/T)\exp(-i\omega_0 t -i\pi/2)$ with $\omega_0=2.417$ fs$^{-1}$ ($780$ nm wavelength) and $T=3.9$ fs (single-cycle pulse with one carrier period $2\pi/\omega_0$ in its FWHM of intensity $|P(t)|^2$). Its spectrum $\hat P_\omega = (T/2)\mbox{tri}[2T(\omega-\omega_0)/\pi]$ (normalized to its peak value) is shown in the inset. [The sinc and tri functions are $\mbox{sinc}(x) = \sin(\pi x)/\pi x$, and $\mbox{tri}(x) = 1-x$ in $0<x<1$, $1+x$ in $-1<x<0$, and 0 for $|x|>1$.] For this pulse $\bar\omega=\omega_0$ and $\sigma^2=0.26$ fs$^{-2}$, yielding $|l|<\bar\omega^2/\sigma^2=22.5$. (b) Fluence profiles (normalized to the peak value in each case) of the RPBs in (\ref{FIXED}) with the indicated values of $|l|$. The vertical dashed lines are placed at $\sqrt{|l|/2}$.
(c) Real part and modulus of $P(t)$ in (\ref{PE}) with $\bar\omega=2.417$ fs$^{-1}$ ($780$ nm wavelength) $\alpha=11.25$ (slightly sub-cycle pulse) and $\Phi=\pi/2$. Its PE spectrum $\hat P_\omega$ (normalized to the peak value) is shown in the inset. For this pulse $\sigma^2=0.26$ fs$^{-2}$, and $|l|<\bar\omega^2/\sigma^2=22.5$, as in (a). (d) Energy density profiles (normalized to the peak value in each case) of the RPBs in (\ref{FIXED}) with the indicated values of $l$. The vertical dashed lines are placed at $\sqrt{|l|/2}$.}
\end{figure}

With the unscaled radius $r$ the bright caustic $\rho_p^2 = |l|/2\bar\omega$ reads $r_p=\sqrt{|l|/2}\, s_{\bar\omega}(z)$, which is the same expression as that of the bright caustic of a monochromatic LG beam of zero radial order with Gaussian width  $s_{\bar\omega}(z)=s_{\bar\omega}\sqrt{1+(z/z_R)^2}$ and Gaussian waist width $s_{\bar\omega} = \sqrt{2z_Rc/\bar \omega}$ of the mean frequency $\bar\omega$. Also, expression (\ref{ISO}) with $\rho_p^2 = |l|/2\bar\omega$ can be more explicitly written, using (\ref{Q}) for the frequency $\bar\omega$, as
\begin{eqnarray}\label{FIXED}
E(r,\phi, z, t') &=& \frac{e^{-i(|l|+1)\psi(z)}e^{-il\phi}}{\sqrt{1+\left(\frac{z}{z_R}\right)^2}}\left(\sqrt{\frac{2}{|l|}}\frac{r}{ s_{\bar\omega}(z)}\right)^{|l|} \nonumber \\
&\times & P\left(t'- \frac{r^2}{2cq(z)}+i\frac{|l|}{2\bar\omega}\right)\,.
\end{eqnarray}
This equation represents a RPB of pulse shape $P(t)$ of maximum energy at $r_p=\sqrt{|l|/2}\, s_{\bar\omega}(z)$ if $P(t)$ is chosen to satisfy $\sigma^2/\bar\omega^2 < 1/|l|$. Otherwise (\ref{FIXED}) is only an unpractical way to specify a RPB of a different pulse shape of maximum energy  such that $\sigma^2/\bar\omega^2 < 1/|l|$ at another location. We stress that (\ref{MAXL}) holds for arbitrarily short pulse forms, not only those having a physically meaningful carrier frequency $\bar\omega$ and envelope $A(t)=P(t)e^{i\bar\omega t}$, i.e., at least one carrier oscillation in the FWHM of $|P(t)|^2$ \cite{BRABEC}. Restriction (\ref{MAXL}) only involves the spectral density $|\hat P_\omega|^2$ and not other characteristics such as the pulse duration. In particular, the topological charge is equally limited for a transform-limited pulse and for a temporally broadened pulse with inhomogeneous spectral phases. Also, we note that the square root of the variance, $\sigma$, is too small to measure the pulse bandwidth, even the half bandwidth, but $\Delta\omega\equiv 2\sigma$ is the so-called Gaussian-equivalent half width \cite{SIEGMAN} ($1/e^2$ decay for a Gaussian-like $|\hat P_\omega|^2$). In terms of the bandwidth $\Delta\omega$, (\ref{MAXL}) reads $|l|< 4\bar\omega^2/\Delta\omega^2$.

Figure \ref{Fig2} is an example that helps to understand the precise meaning of the above result. We try to synthesize RPBs of increasing topological charge but fixed pulse shape at their bright ring, the single-cycle pulse in Fig. \ref{Fig2}(a). For this pulse $|l|$ must be smaller than $\bar\omega^2/\sigma^2=22.5$. The curves in Fig. \ref{Fig2}(b) represent the energy density profiles for different values of $|l|$, and the vertical dashed lines of the same color the radii $r/s_{\bar\omega}(z)=\sqrt{|l|/2}$. With low $|l|$ the energy has a single maximum at the expected radius $r/s_{\bar\omega}(z)=\sqrt{|l|/2}$ (red). A secondary hump begins to emerge with $|l|\simeq 15$ (blue), becomes a maximum with  $|l|=17$ (orange), and the absolute maximum with $|l|=18$ (purple), while the original maximum remains at $r/s_{\bar\omega}(z)=\sqrt{|l|/2}$. With $|l|=23$ (green), the maximum at $r/s_{\bar\omega}(z)=\sqrt{|l|/2}$ turns into a minimum, as predicted, because a new and very weak maximum emerges on its left. This example confirms that inequality (\ref{MAXL}) (obtained from the condition that $r_p$ is a minimum) holds, and also illustrates that this inequality is not sharp in this example, but there is a lower upper bound for $|l|$ (when the secondary maximum becomes the absolute maximum) for this particular pulse.

This observation poses the question of whether exists a pulse shape for which inequality (\ref{MAXL}) is sharp, i. e., $|l|$ can reach the integer part of $\bar\omega^2/ \sigma^2$, and therefore can carry, among all pulses with the same quotient $\bar\omega^2/\sigma^2$, the highest allowed topological charge. We have found that this is the case of the pulses
\begin{equation}\label{PE}
P(t) = \left(\frac{-i\alpha}{\bar\omega t-i\alpha}\right)^{\alpha+1/2}e^{-i\Phi}\,,
\end{equation}
with $\alpha>1/2$ and $\Phi$ an arbitrary phase. Equation (\ref{PE}) is a convenient way to express the commonly used pulses with the power-exponential (PE) or Poisson spectrum \cite{CONTI,PORRAS2,FENG} $\hat P_\omega \propto (\omega/\bar\omega)^{\alpha-1/2}e^{-\alpha\omega/\bar\omega}e^{-i\Phi}$, where the mean frequency appears explicitly. The pulse shape is fully determined by the parameter $\alpha$ and then scaled by $\bar\omega$. The (Gaussian-equivalent) half bandwidth $\Delta\omega=2\sigma \sqrt{2/\alpha}\,\bar\omega$ and half duration $\Delta t =\sqrt{2\alpha}/\bar\omega$ verify $\Delta t \Delta\omega=2$. For the lowest values of $\alpha$ (\ref{PE}) has no a physically meaningful carrier and envelope. With increasing $\alpha$ (\ref{PE}) approaches a Gaussian-enveloped pulse of increasing number of oscillations and Gaussian duration $\Delta t$, $\Phi$ becoming the carrier-envelope phase. For reference, a single-cycle pulse corresponds to $\alpha\simeq 13.75$. With this class of pulses restriction (\ref{MAXL}) reads $|l| < 2\alpha$. For the ``single-cycle" pulse with $\alpha=2.5$ used in \cite{CONTI}, we have $|l|<5$. For the standard single-cycle pulse with $\alpha=13.75$, $|l|<27.5$, i. e., it can carry up to $27$ units of OAM. The energy density of the RPB, normalized to its maximum value at the waist $z=0$, can be calculated to be
\begin{equation}
{\cal E}(r,z) =|D|^2 \left(\frac{2}{|l|}\frac{r^2}{s_{\bar\omega}^2(z)}\right)^{|l|}\left(\frac{\alpha}{\frac{r^2}{s_{\bar\omega}^2(z)}-\frac{|l|}{2} +\alpha}\right)^{2\alpha}\,,
\end{equation}
and is seen to feature for all $|l| < 2\alpha$ one and only one maximum at $r/s_{\bar\omega}(z)=\sqrt{|l|/2}$. For comparison with Fig. \ref{Fig2}(a) and (b), the PE pulse in Fig. \ref{Fig2}(c) is chosen to have $\alpha=11.25$ so that (\ref{MAXL}) yields the same limitation $|l|< 22.5$. The curves and vertical lines in Fig. \ref{Fig2}(d) are the energy profiles with a single maximum at $r/s_{\bar\omega}(z)=\sqrt{|l|/2}$ up to $|l|=22$ (instead of $18$ in preceding example), and with a minimum for $|l|>22$ surrounded by infinite maxima, and thus lacking physical meaning. Being (\ref{MAXL}) a sharp inequality for a class of pulses, this limit cannot be improved for general RPBs.

For our last considerations we note that the RPB (\ref{FIXED}) with the PE pulse (\ref{PE}) in its bright ring $r/s_{\bar\omega}^2(z)=\sqrt{|l|/2}$ if $|l|<2\alpha$, has the peculiarity of having the same PE pulse shape, i. e., the same $\alpha$, at all caustic surfaces, being simply scaled as a whole to the blue-shifted (for $r/s_{\bar\omega}(z)<\sqrt{l/2}$) or red-shifted (for $r/s_{\bar\omega}(z)>\sqrt{l/2}$) mean frequency $\bar\omega(r) = \bar\omega / \{1- \left[r^2/s_{\bar\omega}^2(z) -|l|/2\right]/\alpha\}$. It then makes sense to talk about the pulse shape of the RPB as a whole.

\begin{figure}[b!]
\centering
\includegraphics[width=8.5cm]{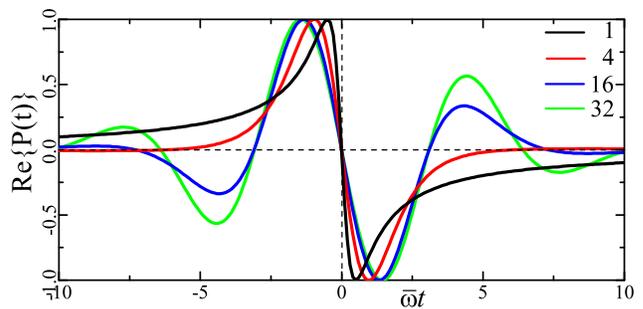}
\caption{\label{Fig3} Shortest $l$-vortex-carrying pulses.}
\end{figure}

An implication of the above results is that there exist minimal wave packets that can carry a $l$-vortex, and in particular a minimal wave packet that can carry a single vortex. Given $|l|$, the Gaussian-equivalent bandwidth of the pulse at the bright ring satisfies $\Delta\omega/\bar\omega < (2/\sqrt{|l|})$, an inequality saturated by the RPB of the PE pulse with $\alpha = |l|/2 +\epsilon$, $\epsilon\rightarrow 0$, of duration $\Delta t=2/\Delta\omega$. We then obtain, for any vortex-carrying pulse
\begin{equation}
\bar\omega \Delta t > \sqrt{|l|}\,.
\end{equation} 
Figure \ref{Fig3} shows shortest $l$-vortex-carrying pulse shapes, that is, (\ref{PE}) with $\alpha=|l|/2 +\epsilon$, for different charges $l$. At any caustic surface the pulse shape is the same with the replacement $\bar\omega\rightarrow \bar\omega(r)$.

We have determined, in conclusion, a limit to the degrees of freedom or eigenvalues $l$ of the OAM of an ultrashort pulse, and a limitation to the duration of a pulse for it to be eigenstate of OAM. It turns out from the literature \cite{ZURCH,HERNANDEZ,GARIEPY,REGO} that the limit (\ref{MAXL}) is close to being reached, and therefore it must be taken into account as a fundamental restriction in future experiments and applications in which shorter pulses with higher OAM are intended to be generated.

The author acknowledges support from Projects of the Spanish Ministerio de Econom\'{\i}a y Competitividad No. MTM2015-63914-P, and No. FIS2017-87360-P.

\end{document}